\documentclass[aps,pra,floatfix,twocolumn,superscriptaddress]{revtex4-2}
\usepackage{graphicx}
\usepackage[english]{babel}
\usepackage{amsmath}
\usepackage{amssymb}
\usepackage{tensor}

\newcommand{\vk}{\mathbf{k}}

\newcommand{\be}{\begin{eqnarray}}
\newcommand{\ee}{\end{eqnarray}}
\newcommand{\p}{\partial}

\newcommand{\dc}{c^{\dagger}}

\def\ket#1{|#1\rangle}

\def\ep#1{\langle #1 \rangle}

\begin{document}

\title{BCS-BEC crossover of atomic Fermi superfluid in a spherical bubble trap}

\author{Yan He}
\affiliation{College of Physics, Sichuan University, Chengdu, Sichuan 610064, China}
\email{heyan$_$ctp@scu.edu.cn}

\author{Hao Guo}
\affiliation{Department of Physics, Southeast University, Jiulonghu Campus, Nanjing 211189, China}
\email{guohao.ph@seu.edu.cn}

\author{Chih-Chun Chien}
\affiliation{Department of physics, University of California, Merced, CA 95343, USA.}
\email{cchien5@ucmerced.edu}

\begin{abstract}
We present a theory of a two-component atomic Fermi gas with tunable attractive contact interactions on a spherical shell going through the Bardeen-Cooper-Schrieffer (BCS) - Bose Einstein condensation (BEC) crossover, inspired by the realizations of spherical bubble traps for ultracold atoms in microgravity. The derivation follows the BCS-Leggett theory to obtain the gap and number equations. The BCS-BEC crossover can be induced by tuning the interaction, and the properly normalized gap and chemical potential exhibit universal behavior regardless of the planar or spherical geometry. Nevertheless, the spherical-shell geometry introduces another way of inducing the crossover by the curvature. The curvature-induced BCS-BEC crossover is made possible by fixing the particle number and interaction strength while shrinking the sphere, causing a reduction to the ratio of the pairing and kinetic energies and pushing the system towards the BCS limit. The saturation of the superfluid density further confirms the ground state is a Fermi superfluid.
\end{abstract}

\maketitle

\section{Introduction}
Ultracold atoms have offered versatile platforms for studying quantum many-body physics with precise controls and broad tunability~\cite{Pethick_book,Ueda_book,Levin_book,Torma_book,Pitaevskii_book,Stoof_book,Gardiner_book,Zhai_book}. While the Bose-Einstein condensation (BEC) has been the foundation behind major research of bosonic atoms~\cite{BEC_collection1,BEC_collection2}, pairing between fermionic atoms introduces the Bardeen-Cooper-Schrieffer (BCS)-BEC crossover that smoothly interpolate the behavior of fermionic and bosonic superfluids~\cite{Inguscio_book,LevinAnnPhys,Zwerger_book,Randeria14,Ohashi20}. On the BCS side of the crossover at zero temperature, the pairing gap is small with respect to the Fermi energy $E_F$ while the chemical potential $\mu$ is near $E_F$. On the BEC side, the gap is comparable to or larger than $E_F$ while $\mu$ becomes negative due to the strong binding of fermions. The mean-field BCS-Leggett theory~\cite{Leggett} captures the main feature of the ground state in the crossover.

Meanwhile, geometry has played an important role in the study of cold atoms. For example, an atomic superfluid in a harmonic trap carries angular momentum by forming vortices~\cite{Madison00}, but an atomic superfluid in a ring-shape trap carries angular momentum by its circulating persistent current~\cite{Eckel14}. 
Another example is the realizations of 2D planar atomic systems, including 2D superfluids~\cite{Desbuquois12,Sobirey21}, 2D BCS-BEC crossover~\cite{Ries15,Enss16}, spin-orbit coupled superfluids~\cite{WuScience16}, phase transitions~\cite{Fletcher15,Murthy15}, scale invariance~\cite{Hung21}, along with many theoretical works~\cite{Botelho06,Bertaina11,HePRL12,Chien2Dboson,HePRA15,Levin2D}. On the other hand, spherical bubble traps for cold atoms have been proposed~\cite{Zobay01,SphericalBECnpj19} and recently realized in microgravity environment, such as the outer space~\cite{Carollo21}. 
Shells of superfluid have also been observed in atomic Mott insulator-superfluid systems~\cite{Folling06} and relevant to neutron stars~\cite{Chamel08}.
While there have been theoretical studies of bosonic superfluid on a spherical shell~\cite{SphericalBECPRL19,SphericalSFPRL20,SphericalBECNJP20,SphericalBECPRA21,SphericalSF21,Rhyno21,Andriati21}, showing enhanced transition temperature, vortices, multi-component mixtures, etc., less references can be found on fermionic superfluid on a spherical shell. 

Here we present an analysis of the BCS-BEC crossover of a Fermi superfluid on a 2D spherical shell at the level of the BCS-Leggett theory. The dispersion of an ideal Fermi gas on a spherical shell already exhibits interesting features~\cite{Cricchio12}, including degeneracy within an angular-momentum level and jumps between adjacent levels. By considering a contact interaction similar to that in nuclear matter~\cite{FetterWalecka}, we obtain a mean-field Hamiltonian describing pairing of the fermions on a spherical shell. Implementing the Bogoliubov transformation~\cite{FetterWalecka}, the gap and number equations on a spherical shell are derived. The solution exhibits the signatures of the BCS-BEC crossover as the attractive interaction increases. When the gap and chemical potential are properly normalized, they exhibit universal behavior that depends only on the interaction but not the curvature, as long as the sphere is large so that the scattering remains a local event.

Nevertheless, the curvature will be shown to influence the Fermi superfluid and induce its own BCS-BEC crossover on a sphere. This is because a bound state always exists in 2D two-body scattering~\cite{Adhikari86,Castin}. In contrast, a two-body bound state in 3D only emerges beyond the unitary point~\cite{ChinRMP10}. The binding energy is determined by the scattering length that quantifies the interaction strength. In experiments, the size of the spherical bubble trap is expected to be tunable with the particle number fixed, so the particle density increases with the curvature. Since the Fermi energy increases with the particle density, the ratio of the pairing energy indicated by the two-body binding energy and the kinetic energy indicated by the Fermi energy decreases as the spherical shell shrinks, thereby pushing the Fermi superfluid towards the BCS limit even when the interaction is fixed. The curvature-induced BCS-BEC crossover is made possible by the 2D nature and the compactness of the spherical bubble trap, and its realization will offer another elegant example of geometric effects on strongly interacting quantum systems. We remark that the topology of a sphere is different from a plane as the Poincare-Hopf theorem~\cite{PHT} states that vector fields on the tangent planes of a sphere must have singularities while those on a plane may have none, and there is a recent study on the XY model on a spherical shell~\cite{Song22}.

The rest of the paper is organized as follows. Sec.~\ref{sec:theory} shows a derivation of the BCS-Leggett theory of fermionic superfluids in the BCS-BEC crossover on a spherical shell. A comparison with the planar case is presented to show the universal behavior. Sec.~\ref{sec:curvature} presents the curvature-induced BCS-BEC crossover as the spherical shell shrinks. Sec.~\ref{sec:implication} discusses theoretical and experimental implications of the BCS-BEC crossover on a spherical shell. Finally, Sec.~\ref{sec:conclusion} concludes our work. Some details and derivations are given in the Appendix.

\section{Effective theory of fermionic superfluid on a spherical shell}\label{sec:theory}
\subsection{Model Hamiltonian}
We consider a two-component atomic Fermi gas with equal mass and population confined in a spherical bubble trap. Assuming the shell is thin, the gas thus lives on the surface of a sphere. For a free Fermi gas confined on a spherical shell, the energy dispersion is given by~\cite{Cricchio12}
\be
\epsilon_l=\frac{\hbar^2}{2mR^2}l(l+1),\quad l=0,1,\cdots.
\ee
Here $m$ is the mass of the atoms and $R$ is the radius of the sphere. In the following, we will set $\hbar=1$ and $k_B=1$. $l$ is the quantum number of the orbital angular momentum. For a fixed $l$, the magnetic quantum number takes the values $m_z=-l,\cdots, l$, and $\sigma=\uparrow,\downarrow$ labels the two components. Therefore, there are $2(2l+1)$ degenerate states for the level labeled by $l$. 

After including a two-body interaction term modeling atomic scattering, the Hamiltonian in the grand-canonical ensemble is $H=H_K+H_I$, where $H_K=\sum_{l,m,\sigma} (\epsilon_l-\mu)\dc_{lm\sigma} c_{lm\sigma}$ and
\be
H_I&=&\sum'_{l_1,m_1,\cdots}V_{1234}\dc_{l_1 m_1\sigma_1}\dc_{l_2 m_2\sigma_2} c_{l_3 m_3\sigma_3} c_{l_4 m_4\sigma_4}.
\ee
Here $V_{1234}=\ep{l_1, m_1; l_2, m_2|V|l_3, m_3; l_4, m_4}$ and $\dc_{lm\sigma}$ ($c_{lm\sigma}$) is the fermion creation (annihilation) operator. We also assume equal populations of the two component, so $\mu_{\sigma}=\mu$. Assuming the two-body interaction is rotational invariant, then non-vanishing matrix elements only occur if the magnetic quantum numbers satisfy $m_1+m_2=m_3+m_4$, as indicated by the prime above the $\sum$. For atomic gases, the interactions are usually tunable via Feshbach resonance by an external magnetic field~\cite{Pethick_book,BEC_collection1,ChinRMP10}. Conventional superconductors are due to phonon mediated interactions~\cite{Tinkham}. In principle, one may formulate mediated interactions on a sphere. The expressions may be more complicated than those presented here, and it may be challenging to tune those mediated interactions through the BCS-BEC crossover.

In the conventional BCS theory, one only considers two-body scattering from $\ket{\pm\vk}$ to $\ket{\pm\vk'}$, forming Cooper pairs with zero total momentum~\cite{Pethick_book,FetterWalecka}. Inspired by such a simplification, we also focus on the initial and final states on the spherical shell that can be coupled into $\ket{L=0,M=0}$ with spin singlet and ignore other scattering processes. The approximation then leads us to the reduced interaction Hamiltonian
$H_I=\sum_{l_1,m_1,l_2,m_2}V_{12}
\dc_{l_1 m_1\uparrow}\dc_{l_1,-m_1\downarrow} c_{l_2 m_2\uparrow} c_{l_2,-m_2\downarrow}$.
Here $V_{12}=\ep{l_1, m_1; l_1, -m_1|V|l_2, m_2; l_2, -m_2}$.
The coupling among the angular-momentum states gives
$\ket{l, m; l, -m}=\sum_{L=0}^{2l}\ket{l, l; L, 0}\ep{l,l; L, 0|l, m; l, -m}$.
Here we only keep the $L=0$ state and use the fact that
$\ep{l, l; L, 0|l, m; l, -m}=(-1)^{l-m}/\sqrt{2l+1}$.
The interaction Hamiltonian then becomes $H_I=\sum_{l_1,m_1,l_2,m_2}V_{12,0}
\dc_{l_1 m_1\uparrow}\dc_{l_1,-m_1\downarrow} c_{l_2 m_2\uparrow} c_{l_2,-m_2\downarrow}$. 
Here $V_{12,0}=\ep{l_1,l_1; 0, 0|V|l_2, l_2; 0, 0}\frac{(-1)^{l_1-m_1}(-1)^{l_2-m_2}}{\sqrt{(2l_1+1)(2l_2+1)}}$. The factor $(-1)^{l_1-m_1}(-1)^{l_2-m_2}$ inside $V_{12,0}$ can be removed by a canonical transformation, given by
$c_{lm}\to c_{lm}$ and $c_{l-m}\to(-1)^{l-m}c_{l-m}$.
After those calculations, 
the form of $H_I$ is now suitable for a general mean-field approximation similar to the BCS theory.

\subsection{BCS theory on a spherical shell}
Following the BCS approximation, we make the substitutions
$c_{l m\uparrow} c_{l,-m\downarrow}\to \ep{c_{l m\uparrow} c_{l,-m\downarrow}}$ and
$\dc_{l m\uparrow}\dc_{l,-m\downarrow}\to \ep{\dc_{l m\uparrow}\dc_{l,-m\downarrow}}$
in the interaction Hamiltonian and keep only up to the quadratic terms. This leads to the BCS Hamiltonian
\be
H_{BCS}&=&H_K+\sum_{l,m}(-\Delta_l\dc_{l m\uparrow}\dc_{l,-m\downarrow}-\Delta_l c_{l m\uparrow}c_{l,-m\downarrow}).
\ee
Here the gap function is given by
\be
&&\Delta_j=-\frac{1}{(2j+1)^{1/2}}\sum_{l,m}V_{jl}\frac{1}{(2l+1)^{1/2}}\ep{c_{l m\uparrow}c_{l,-m\downarrow}}
\ee
and $V_{jl}=\ep{j,j;0,0|V|l,l;0,0}$.
The BCS Hamiltonian can be diagonalized by the Bogoliubov transformation~\cite{FetterWalecka} with
\be
&&c_{lm\uparrow}=u_{l}\alpha_{lm}-v_{l}\beta_{lm},~~ \dc_{l,-m\downarrow}=v_{l}\alpha_{lm}+u_{l}\beta_{lm}.
\ee
The coefficients are given by
$u_{l}^2=\frac12(1+\frac{\xi_{l}}{E_{l}})$ and $v_{l}^2=\frac12(1-\frac{\xi_{l}}{E_{l}})$,
where $\xi_{l}=\epsilon_{l}-\mu$ and $E_{l}=\sqrt{\xi_{l}^2+\Delta_{l}^2}$.
The diagonalized Hamiltonian has the form
\be
H_{BCS}=\sum_{lm}(\xi_{l}-E_{l})+\sum_{lm}E_l (\alpha^{\dag}_{lm}\alpha_{lm}+\beta^{\dag}_{lm}\beta_{lm}).
\ee

In terms of the Bogoliubov transformation, the gap function becomes
\be
\Delta_j=\frac{-1}{(2j+1)^{1/2}}\sum_{l}V_{jl}(2l+1)^{1/2}u_{l}v_{l}[1-2f(E_{l})].
\ee
Here $f(x)=1/[\exp(x/T)+1]$ is the Fermi distribution function. 
We will further approximate the matrix element $V_{jl}$ before solving the gap equation. 
Meanwhile, the number equation can be derived from $n=\sum_{l,m,\sigma}\langle\dc_{lm\sigma} c_{lm\sigma}\rangle$. Explicitly,
\be
n=\sum_{l}(2l+1)\Big(1-\frac{\xi_{l}}{E_{l}}+2\frac{\xi_{l}}{E_{l}}f(E_{l})\Big).
\ee
Solving the gap and number equations gives us $\Delta$ and $\mu$ of the Fermi gas.

\begin{figure}
\centering
\includegraphics[width=\columnwidth]{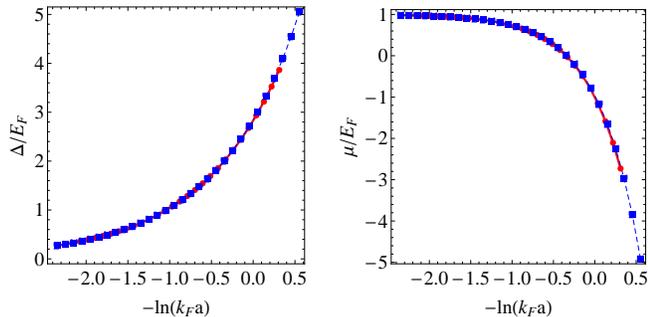}
\caption{Universal behavior of interaction-induced BCS-BEC crossover: The normalized gap (left) and chemical potential (right) as a function of $-\ln(k_F a)$ on a 2D plane according to Eq.~\eqref{eq:gn_plane} (red circles) and on the shell of a unit sphere according to Eq.~\eqref{eq:gn_sphere} (blue squares) at zero temperature.}
\label{delta}
\end{figure}

\subsection{BCS-BEC crossover on a spherical shell}
We begin with a brief review of the mean-field description of the BCS-BEC crossover on a 2D plane, following Refs. \cite{Botelho06,Levin2D}. To handle the bound state from the 2D two-body scattering, a regularization introduces a binding energy $\epsilon_b=-\hbar^2/(ma^2)$, where $a$ is the 2D two-body $s$-wave scattering length. Combining the binding energy with the renormalization of the contact interaction, the coupling constant is expressed in terms of the scattering length via
$\frac 1g=\int\frac{d^2k}{(2\pi)^2}\frac{1}{2\epsilon_\vk+|\epsilon_b|}$.
Here $\epsilon_\vk=\hbar^2 k^2/(2m)$ is the free-fermion dispersion.
The gap and number equations can be simplified to
\be \label{eq:gn_plane}
&&\int\frac{d^2k}{(2\pi)^2}\Big[\frac{1-f(E_\vk)}{2E_\vk}-\frac{1}{2\epsilon_\vk+|\epsilon_b|}\Big]=0. \\
&& n=\int \frac{d^2k}{(2\pi)^2}\Big(1-\frac{\xi_{k}}{E_{k}}+2\frac{\xi_{k}}{E_{k}}f(E_{k})\Big). \nonumber
\ee
Solving the equations gives $\Delta$ and $\mu$ once the values of $a$ and $T$ are given.

In a previous study of bosonic atoms in a spherical-shell potential~\cite{SphericalBECPRL19}, a contact interaction has been implemented. To simplify the BCS theory on a spherical shell, we also implement an approximation of the matrix element $V_{jl}$ by considering only a short-range attractive interaction. A choice is a two-body contact interaction of the form
$V=-g\delta(1-\cos\theta_{12})$,
where $\cos\theta_{12}=\cos\theta_1\cos\theta_2+\sin\theta_1\sin\theta_2\cos(\phi_1-\phi_2)$. As one will see shortly, the choice renders a constant gap function. The coupling constant $g$ will be related to the 2D scattering length in a discussion later. After using a generalization of the Wigner-Eckart theorem~\cite{FetterWalecka} as explained in Appendix~\ref{app:Vsimp},
the matrix element becomes
$V_{jl}=-g\sqrt{(2j+1)(2l+1)}$. We remark that the contact potential has infinitesimal interaction range. When we expand the contact potential by spherical harmonics, the expansion coefficients are similar in magnitude for all angular momentum $l$, leading to the simplified expression of the matrix elements. If a different interaction potential is considered, the dominant contributions may come from those with small $l$, and the strength decays as $l$ increases.

The gap equation is then reduced to
$\Delta_j=g\sum_{l}(2l+1)\frac{\Delta_l}{2E_l}[1-2f(E_{l})]$.
Since the right hand side does not depend on $j$, we conclude that $\Delta$ does not depend on $j$ explicitly. Hence, the gap equation reduces to
$\frac{1}{g}=\sum_{l}\frac{2l+1}{2E_l}[1-2f(E_{l})]$.
Since $E_l\propto l(l+1)$, the dominant terms in the summation will behave like
$\sum_{l}\frac{2l+1}{2E_l}\sim \sum_{l}\frac{2l+1}{l(l+1)}\to \infty$ due to the contact-interaction approximation. A systematic renormalization scheme, similar to the one in flat space, can be applied to render meaningful physical results.

Following the planar case, the regularization on a 2D spherical shell can be modified as
$\frac 1g=\int dl\, \frac{2l+1}{2\epsilon_l+|\epsilon_b|}$.
We assume $\epsilon_b=-\hbar^2/(ma^2)$ due to its localized nature. The two-body scattering length can be measured experimentally to characterize the interaction strength~\cite{Pethick_book,ChinRMP10}.
After the regularization, we obtain the gap and number equations as
\be \label{eq:gn_sphere}
&&\int dl\,(2l+1)\Big[\frac{1-2f(E_{l})}{2E_l}-\frac{1}{2\epsilon_l+|\epsilon_b|}\Big]=0, \label{eq:gap1}\\
&&n=\frac{1}{4\pi R^2}\int dl\,(2l+1)\Big(1-\frac{\xi_{l}}{E_{l}}+2\frac{\xi_{l}}{E_{l}}f(E_{l})\Big).\nonumber
\ee
We mention there is another regularization scheme summarized in Appendix~\ref{app:ren} that produces qualitatively the same results. Moreover, we have approximate the summations by integrals, and a comparison in Appendix~\ref{app:IntVsSum} shows that there is no observable difference between the results from the summations and the integrals for reasonably large $l$.

Numerical results of the BCS-BEC crossover on a 2D spherical shell at zero temperature are shown in Fig.~\ref{delta}, along with the results of the 2D planar case.
We plot $\Delta$ and $\mu$ as a function of $-\ln(k_F a)$ for both cases. 
For the 2D planar case, $E_F$ and $k_F$ are the Fermi energy and Fermi momentum of a noninteracting Fermi gas with the same density. 
For the spherical-shell case, we take $E_F$ and $k_F=\sqrt{2mE_F}$ from a noninteracting Fermi gas with the same total particle number. Assuming the largest occupied shell has angular momentum $L_m$ for a free Fermi gas, the total particle number is $N=2L_m(L_m+1)$, so $E_F=\frac{L_m(L_m+1)}{2mR^2}$ and $n=N/(4\pi R^2)$. As $-ln(k_F a)$ increases, the gap increases while the chemical potential decreases, showing the signature of the BCS-BEC crossover. While the BCS-BEC crossover is not a sharp transition, the crossover may be identified as the regime where $\mu$ changes sign. Since the scattering length reflects the effective interactions between the fermions, Fig.~\ref{delta} shows the interaction induced BCS-BCS crossover in two different 2D geometries.

Importantly, when normalized according to their respective intrinsic quantities like $E_F$ and $k_F$, the results of the 2D plane are indistinguishable from those of the spherical shell. This is because the pairing from the contact interaction is a local property of the Fermi gas. As a consequence, properly normalized quantities reflect the same local behavior from the mean-field theory and fail to differentiate the global geometry. 
For the spherical case, taking different values of $N$, $L_m$, and $R$ produces the same universal results of $\Delta/E_F$ and $\mu/E_F$. 
The universal behavior can also be confirmed by the resemblance of the equations of state, Eqs.~\eqref{eq:gn_plane} and \eqref{eq:gn_sphere}, when written in the normalized quantities. The details can be found in Appendix~\ref{app:Uni}. We remark that by using dimensionless quantities such as $\Delta/E_F$ and $\mu/E_F$, the theory can be applied to systems with different species of atoms and different sizes or numbers to extract universal behavior.

\begin{figure}
\centering
\includegraphics[width=\columnwidth]{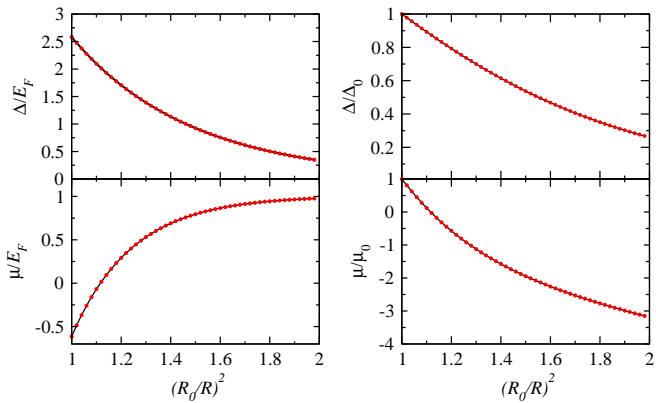}
\caption{Curvature-induced BCS-BEC crossover on a spherical shell at zero temperature: (Top panels) Gap as a function of $(R_0/R)^2$ normalized to $E_F$ (left) and $\Delta_0$ (right). (Bottom panels) Chemical potential as a function of $(R_0/R)^2$ normalized to $E_F$ (left) and $\mu_0$ (right). $\Delta_0$ and $\mu_0$ on the right panels are the gap and chemical potential of $N=220$ fermions on the shell of a reference sphere with radius $R_0$ satisfying $-\ln(k_F a)=-0.1$. Then $a$ and $N$ are fixed while $(R_0/R)$ varies.
}
\label{del-R}
\end{figure}

\section{Curvature induced BEC-BCS crossover}\label{sec:curvature} 
The universal behavior of
Fig.~\ref{delta} may lead to a false impression that the curvature of the sphere, $1/R^2$, does not play a significant role. However, one may envision that the radius of the spherical bubble trap is tunable and consider a different scenario where the particle number, not the local density, is conserved and compare the physical quantities with different curvatures but the same interaction strength. As the curvature increases, the surface area shrinks and the local particle density increases if the total particle number is fixed. Since $E_F$ increases with the density, it is tempting to claim that the gap will increase with the curvature if $\Delta/E_F$ is roughly the same. A careful analysis, however, reveals the opposite and establish a BEC-BCS crossover induced by the curvature.

To demonstrate the curvature effects, we plot $\mu$ and $\Delta$ as functions of the curvature with fixed particle number and scattering length. As the radius of the sphere shrinks from $R_0$, one can see that the gap become smaller with respect to the gap at $R_0$. The reason is that as $1/R^2$ increases, the Fermi energy becomes larger. Meanwhile, the effective two-body binding energy is fixed by the scattering length, which is controlled by an external magnetic field. The ratio $|\epsilon_b|/E_F$ thus decreases with the curvature, resulting in a situation where the kinetic energy dominates the pairing energy and thereby driving the system into the BCS limit as the radius of the spherical shell shrinks. Again, a sign change of $\mu$ indicates the occurrence of the BCS-BEC crossover.

Thus, there are two ways to sweep a Fermi superfluid across the BCS-BEC crossover on a spherical shell, one by tuning the interaction and the other by tuning the geometry. The first one is an analogue of the 2D planar case, where the particle density is fixed and the scattering length is tuned via magnetic or optical means. The second one requires a compact 2D geometry, where the particle number and scattering length are fixed but the ratio between the two-body binding energy and Fermi energy is tuned by the geometry. We remark that the latter is possible in 2D because the two-body binding energy is always present~\cite{Adhikari86,Castin}, different from the general 3D case where the binding energy is finite only on the BEC side. Therefore, shrinking a 3D bulk Fermi superfluid cannot push the system to the BCS regime.
We also caution that the calculation of the two-body scattering length assumes the system is locally flat, and the assumption breaks down when the curvature is too large, or when $R\sim O(a)$. Moreover, we note that the BCS-Leggett theory does not take into account the induced interaction~\cite{Pethick_book} and Hartree-Fock energy~\cite{Miahila11}. The former reduces the transition temperature by a factor in the 3D case, and the latter shifts the chemical potential. As a first attempt to develop the BCS-Leggett theory of Fermi superfluids on a spherical shell, we leave those effects for future, more refined studies.

Furthermore, we evaluate the superfluid density given by
\be
n_s=n-\frac{1}{4\pi R^2}\int dl\,2(2l+1)\frac{l(l+1)}{2mR^2}\Big[-\frac{d f(E_l)}{d E_l}\Big].
\label{eq:ns}
\ee
A derivation based on linear response theory is shown in Appendix~\ref{app:SFder}.
As $T\to0$, $-\frac{df(x)}{dx}$ approaches the delta function. Since $E_l$ is positive, the delta function can never be satisfied. Therefore, $n_s/n=1$ at $T=0$ across the whole interaction-induced BCS-BEC crossover, so the ground state is indeed a Fermi superfluid. Nevertheless, in the curvature-induced crossover, the density increases with the curvature because the total particle number is fixed, leading to an interesting scenario where $n_s$ increases while $\Delta$ decreases with the curvature according to the upper-right panel of Fig.~\ref{del-R}. The disparity of the dependence of $n_s$ and $\Delta$ on the curvature has its root in that $\Delta$ is associated with thermodynamics while $n_s$ is from linear response to perturbations.

\section{Implications}\label{sec:implication}
\begin{figure}
\centering
\includegraphics[width=\columnwidth]{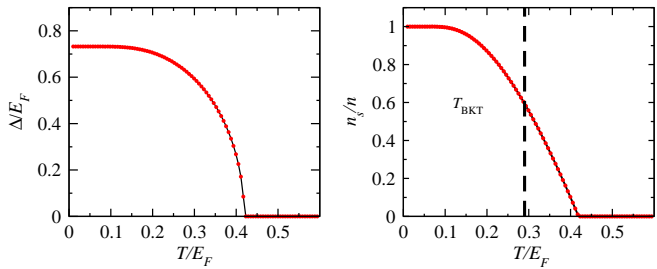}
\caption{Gap (left) and superfluid  density (right) as functions of $T/E_F$ of a Fermi superfluid on the shell of a unit sphere. Here $-\ln(k_F a)=-1.35$ and $N=220$. $T^*/E_F\approx 0.42$ is where $\Delta$ approaches zero. The vertical dashed line indicates $T_{BKT}/E_F\approx 0.29$, above which $n_s$ is expected to drop to zero.}
\label{ns}
\end{figure}

After characterizing the ground-state properties of atomic Fermi superfluids on a spherical shell, we investigate the mean-field theory away from zero temperature by solving the gap and number equations at finite temperatures. 
In Figure~\ref{ns}, we show $\Delta$ and $n_s$ as functions of $T$. The mean-field transition temperature $T^*$ is the point above which $\Delta$ vanishes. As the system moves towards the BEC limit, $T^*$ increases without bound and indicates the pairing energy scale.
The 2D Berezinskii–Kosterlitz–Thouless (BKT) transition~\cite{Berezinskii71,Berezinskii72,KT73} temperature separates the superfluid and normal phase, which may be estimated by 
\be
\dfrac{k_B T_{BKT}}{\hbar^2 n_s(T_{BKT})/m}=\dfrac{\pi}{2}. 
\ee
For the case shown in Figure \ref{ns}, $T_{BKT}$ is below $T^*$, so the BKT transition will preempt the mean-field transition and cause $n_s$ to jump to zero.

\begin{figure}
\centering
\includegraphics[width=0.8\columnwidth]{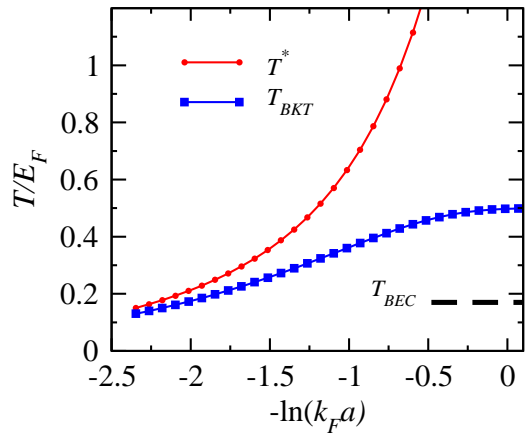}
\caption{$T^*/E_F$ and $T_{BKT}/E_F$ as a function of $-\ln(k_F a)$. Here $N=220$. The dashed line indicates the ideal $T_{BEC}/E_F$ in the BEC regime.}
\label{Tc}
\end{figure}

The BCS-Leggett theory only provides a qualitative description of the crossover at finite temperatures. It has been shown~\cite{Inguscio_book,LevinAnnPhys,Zwerger_book,Randeria14,Ohashi20} that the preformed pairs, which are the analogue of thermal bosons in a Bose gas, lead to a substantially lower $T_c$ on the BEC side. There have been studies of 2D planar Fermi superfluids that include pairing fluctuations~\cite{HePRA15,Levin2D,Enss16} and studies of 2D BKT transition in Fermi superfluids with fluctuation effects~\cite{Botelho06,Tempere09}. In the BEC limit, the tightly bound pairs resemble composite bosons. The BEC temperature of a noninteracting Bose gas on a sphere is given by~\cite{SphericalBECPRL19} 
\be
k_B T_{BEC}=\frac{\frac{2\pi\hbar^2}{m_B} n_B}{\frac{\hbar^2}{m_BR^2 k_B T_{BEC}}-\ln(e^{\hbar^2/m_BR^2k_B T_{BEC}}-1)}.
\ee
By setting $m_B=2m$ and $n_B=n/2$ for the composite bosons, we estimate the ideal BEC temperature of the composite bosons in the BEC regime. For 3D Fermi superfluids in the BEC limit, pairing fluctuations via the Nozieres–Schmitt-Rink and other methods show that the transition temperature approaches $T_{BEC}$~\cite{Randeria14,LevinAnnPhys,Pascucci21} due to the composite bosons, and the same mechanism should apply to 2D systems. Fig.~\ref{Tc} shows $T^*$, $T_{BKT}$, and $T_{BEC}$ as functions of the interaction strength. On the BCS and BEC sides, the BKT and BEC temperatures limits where superfluid and condensate can be observed, respectively, while the mean-field $T^*$ shows where pairing energy enters the excitation spectrum. The three temperatures provide upper bounds for the transition temperatures, and
a full treatment of the finite-temperature BCS-BEC crossover on a spherical shell will be worth another publication. We remark that the BCS, BKT, and BEC transitions are defined in the thermodynamic limit. For a finite system, the transitions will lose the sharpness due to finite-size effect.

\begin{figure}
\centering
\includegraphics[width=\columnwidth]{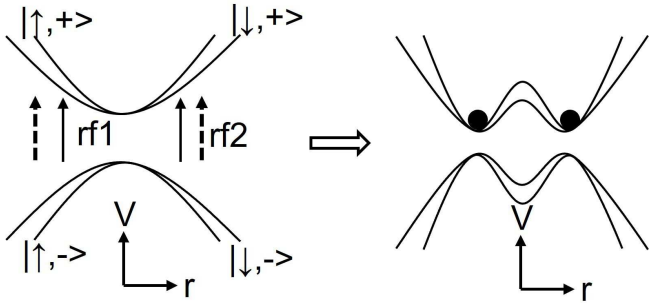}
\caption{Illustration of a bubble trap for two-component atomic Fermi gases. By using radio-frequency (rf) photons (rf1 and rf2) to couple two pairs of hyperfine states $\left |\uparrow,\pm \right\rangle$ and $\left|\downarrow,\pm \right\rangle$ in a harmonic trap with a magnetic field, the resulting minimum in the upper-branch of the potential for the two-component Fermi gas (indicated by the black dots) corresponds to a spherical shell in 3D.}
\label{trap}
\end{figure}

Trapping of single-species atomic bosons in a spherical shell has been achieved by having three hyperfine states in a spherical harmonic potential with energy levels split by a magnetic field~\cite{Carollo21}. A radio-frequency (rf) excitation only couples the hyperfine states at a given radius due to the inhomogeneity from the harmonic trap and inverts the potential inside. The combination of the potentials for the dressed states thus resembles a shell at specific radius determined by the harmonic trap, magnetic field, and rf excitation. Therefore, the shell size can be tuned by the background harmonic potential or radio frequency. The method should in principle work for fermionic atoms. However, to have two components of fermions in a spherical shell, more hyperfine states with selected rf excitations among them may be needed. High-component atomic Fermi gases have been realized~\cite{Taie10}, and they may be suitable for the realization of two-component Fermi gases in a spherical shell in the future. Figure~\ref{trap} illustrate the setup for a spherical bubble trap for two-component fermionic atoms.

In this work, a contact interaction has been implemented to model the atomic collisions. Numerical calculations have shown that short-range interactions describes atomic interactions reasonably~\cite{Carlson11}, and one may consider finite-range interactions to include corrections beyond the low-energy limit~\cite{ParishPRB05,TononiNJP18}. We also remark that when the spherical-shell size shrinks too much, the continually increasing density will lead to strong three-body loss~\cite{ChinRMP10}. Moreover, finite-range corrections to the interaction may become observable when atoms are closely packed. Therefore, while the increasing of the chemical potential and decreasing of the gap function due to the curvature-induced BCS-BEC crossover should be observable as the radius decreases, the behavior of the system may start to deviate from the mean-field description as the trap size gets too small. Ref.~\cite{Carollo21} shows $10^4$ atoms confined in a shell potential with linear size of the order of 100 $\mu$m. For a spherical shell of similar size, this gives a surface density of about $10^{11}$/m$^2$. If a Feshbach resonance is sufficiently far away from others, the scattering length almost covers the range $(-\infty,\infty)$~\cite{ChinRMP10}. When finite thickness of the atomic cloud is considered, the system gradually deviates from the 2D case and eventually becomes 3D when the thickness is comparable to the scattering length. 
For the curvature-induced BCS-BEC crossover, the interaction is assumed to be fixed while the radius changes. Thus, the thickness of the cloud should be roughly the same to keep the scattering properties fixed.

We remark the spherical-shell trap is not the only way for realizing compact 2D geometries for cold atoms. One may, for example, confine planar 2D atomic gases in a finite regime. The distortion of the condensate wavefunction near the boundary is determined by the healing length~\cite{FetterWalecka}, which depends on the interaction and density. One may also envision wrapping a rectangle into the surface of a torus to eliminate boundary effects. The 2D torus has two different principal curvatures while the sphere has the same curvature everywhere. It has been shown~\cite{Khanna14} that the Ginzburg-Landau theory on the surface of a torus exhibits size-dependent transition temperature. However, a torus-surface trap for cold-atoms may be more challenging.

\section{Conclusion}\label{sec:conclusion}
We have presented a generalization of the BCS-Leggett theory of atomic Fermi superfluids on a 2D spherical shell undergoing the BCS-BEC crossover, relevant to future experiments using spherical bubble traps in microgravity. Although the highly degenerate levels and jumps between the levels of an ideal Fermi gas on a spherical shell makes the spectrum different from that on a 2D plane, the pairing gap and chemical potential of a Fermi superfluid after proper normalization exhibit universal behavior transcending the underlying geometries. Therefore, the conventional interaction-induced BCS-BEC crossover of Fermi superfluid is also present on a spherical shell. Nevertheless, the spherical geometry introduces the curvature-induced BCS-BEC crossover by fixing the interaction strength and particle number while reducing the size of the spherical shell. The latter type of crossover is due to a suppression of the ratio between the pairing and kinetic energies by the curvature.
Our work paves the way towards a systematic investigation of Fermi superfluids with compact geometries, exemplified by the spherical bubble traps.

\begin{acknowledgements} 
Y. H. was supported by the Natural Science Foundation  of  China  under  Grant  No.   11874272  and  Science Specialty  Program  of  Sichuan  University  under  Grant No. 2020SCUNL210.   C.  C.  C.  was  supported  by  the National  Science  Foundation  under  Grant  No.    PHY-2011360.
\end{acknowledgements}

\appendix
\section{Calculation of $V$}\label{app:Vsimp}
The two-body contact interaction $V=-g\delta(1-\cos\theta_{12})$ allows an expansion by the Legendre polynomials and spherical harmonic functions as
\be
V&=&-g\sum_L (2L+1)P_L(\cos\theta_{12}) \nonumber \\ 
&=&-4\pi g\sum_{LM}(-1)^M Y_{LM}(\theta_1,\phi_1)Y_{L,-M}(\theta_2,\phi_2).
\ee
Here we treat $Y_{LM}$ as an irreducible tensor operator, so the summation
\be
\sum_{LM}(-1)^M Y_{LM}(\theta_1,\phi_1)Y_{L,-M}(\theta_2,\phi_2)
\ee
is actually a tensor product of two irreducible tensor operators, which results in a scalar operator.

According to a more general version of the Wigner-Eckart theorem shown in Eq. (B.33) of Ref.~\cite{FetterWalecka}, we find that
\be
&&\ep{l_1 l_1 0 0|V|l_2 l_2 0 0}=\nonumber\\
&&-4\pi g\sum_L(-1)^{l_1+l_2}\left\{
  \begin{array}{ccc}
    0 & l_1 & l_1 \\
    L & l_2 & l_2
  \end{array}
\right\}\ep{l_1\|Y_L\|l_2}^2.
\ee
Here the 6j symbol is given by
\be
\left\{
  \begin{array}{ccc}
    0 & l_1 & l_1 \\
    L & l_2 & l_2
  \end{array}
\right\}=(-1)^{l_1+l_2}\frac{1}{\sqrt{(2l_1+1)(2l_2+1)}}
\ee
and reduced matrix element is
\be
&&\ep{l_1\|Y_L\|l_2}=\nonumber\\
&&(-1)^{l_1}\sqrt{\frac{(2l_1+1)(2L+1)(2l_2+1)}{4\pi}}\left(
  \begin{array}{ccc}
    l_1 & L & l_2 \\
    0 & 0 & 0
  \end{array}
\right).
\ee
In the above equation, the large parentheses denote the 3j symbol, not to be confused with the 6j symbol. After collecting all the results, we find that
\be
\ep{l_1 l_1 0 0|V|l_2 l_2 0 0}
&=&-g\sum_L\sqrt{(2l_1+1)(2l_2+1)}\nonumber\\
& &\times(2L+1)\left(
  \begin{array}{ccc}
    l_1 & L & l_2 \\
    0 & 0 & 0
  \end{array}
\right)^2.
\ee
Moreover, the normalization condition of the Clebsch–Gordan (CG) coefficients lead to
\be
& &\sum_L(2L+1)\left(
  \begin{array}{ccc}
    l_1 & L & l_2 \\
    0 & 0 & 0
  \end{array}
\right)^2=\sum_L\ep{l_1 l_2 00|l_1 l_2 L 0}^2=1. \nonumber \\
&&
\ee
After some algebra, the matrix element takes the form 
\be
\ep{l_1 l_1 0 0|V|l_2 l_2 0 0}
=-g\sqrt{(2l_1+1)(2l_2+1)}.
\ee

\section{Alternative renormalization scheme}\label{app:ren}
There is another way of regularizing the gap equation of a Fermi superfluid on a 2D plane. This has been shown in Eq. (12) of Ref.~\cite{Castin} as follows.
\be
\frac 1g=\lim_{q\to0}\Big[-\frac{m}{2\pi}\ln(\frac{aq e^\gamma}{2})
-\int\frac{d^2k}{(2\pi)^2}\mathcal{P}\frac1{2(\epsilon_{\mathbf{q}}-\epsilon_\vk)}\Big].
\ee
Here $\gamma$ is the Euler constant and $\mathcal{P}$ denotes the Cauchy principle value.
Combining with the gap equation of the 2D Fermi superfluid, one finds
\be
-\frac{m}{2\pi}\ln(\frac{aq e^\gamma}{2})&=&\int\frac{d^2k}{(2\pi)^2}\frac{1-f(E_\vk)}{2E_\vk}
+ \nonumber \\
&&\int\frac{d^2k}{(2\pi)^2}\mathcal{P}\frac1{2(\epsilon_{\mathbf{q}}-\epsilon_\vk)}.
\ee
The drawback of this method, however, is that one has to assume an infrared (IR) cutoff $q$. We have verified that this alternative regularization gives qualitatively the same results as those presented in the main text.

\begin{figure}
\centering
\includegraphics[width=\columnwidth]{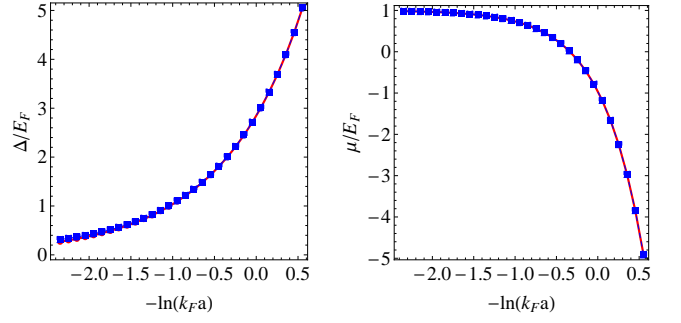}
\caption{Gap and chemical potential as functions of $-\ln(k_F a)$ from Eq.~(\ref{eq:gap1}) (red circles) using the integral and from Eq.~(\ref{eq:gap2}) (blue squares) using the summation. Here $N=220$ and $k_F R=10.5$.}
\label{cmp}
\end{figure}

\section{Universal behavior}\label{app:Uni}
The universal behavior of the gap and chemical potential in the interaction-induced BCS-BEC crossover comes from the equations of state. We let $b=-\ln(k_F a)$, which is equivalent to $a=e^{-b}/k_F$. For the 2D-plane case, $E_F=k_F^2$ when $\hbar=1$ and $2m=1$. The particle number per unit area of a noninteracting Fermi gas is given by $N=2(\pi k_F^2)/(2\pi)^2$, or $n/k_F^2=1/(2\pi)$. Therefore, the gap and number equations of Fermi superfluid on a 2D plane can be written as
\be \label{eq:gnp_norm}
&&\int dy\Big[\frac{1-f(E_k/E_F)}{2E_k/E_F}-\frac{1}{2y^2+2e^{2b}}\Big]=0. \\
&& \frac{1}{2\pi}=\int\frac{dy y}{2\pi}
\Big(1-\frac{\xi_{k}/E_F}{E_{k}/E_F}+2\frac{\xi_{k}/E_F}{E_{k}/E_F}f(E_{k}/E_F)\Big). \nonumber
\ee
Here $y=k/k_F$ and only $\Delta/E_F$ and $\mu/E_F$ show up in $E_k/E_F$ and $\xi_k/E_F$.

Meanwhile, for a noninteracting Fermi gas on a sphere filled up to the angular-momentum state $L_m$, we have $n=N/(4\pi R^2)$, $N=2L_m(L_m+1)$, $E_F=N/(2R^2)$, and $E_F=k_F^2$ with $2m=1$. Again, let $b=-\ln(k_F a)$. The equations of state of Fermi superfluid on a spherical shell thus becomes
\be \label{eq:gns_norm}
&&\int dz\Big(z+\frac{1}{2L_m}\Big)\Big[\frac{1-f(E_l/E_F)}{2E_l/E_F}-\frac{1}{2z^2+2e^{2b}}\Big]=0. \\
&& \frac{1}{2\pi}=\int \frac{dz}{2\pi} \Big(z+\frac{1}{2L_m}\Big)\Big(1-\frac{\xi_{l}/E_F}{E_{l}/E_F}+2\frac{\xi_{l}/E_F}{E_{l}/E_F}f(E_{l}/E_F)\Big). \nonumber
\ee
Here $z=l/L_m$ and only $\Delta/E_F$ and $\mu/E_F$ show up in $E_l/E_F$ and $\xi_l/E_F$. When $L_m\gg 1$, which is usually the case in many-body systems, the two sets of equations of state, Eqs.~\eqref{eq:gnp_norm} and \eqref{eq:gns_norm}, are identical and give the universal behavior of the normalized gap and chemical potential.

\section{Integral vs. summation in the equations}\label{app:IntVsSum}
Here we compare the results from the gap and number equations using summation over the angular momentum versus the approximation using integration. The equations with explicit summations are 
\be
&&\sum_{l=0}^{L_M}(2l+1)\Big[\frac{1-2f(E_{l})}{2E_l}-\frac{1}{2\epsilon_l+|\epsilon_b|}\Big]=0, \label{eq:gap2}\\
&&N=\sum_{l=0}^{L_M}(2l+1)\Big(1-\frac{\xi_{l}}{E_{l}}+2\frac{\xi_{l}}{E_{l}}f(E_{l})\Big).\nonumber
\ee
Here $L_M$ is some cutoff level, which is much larger than the highest occupied shell $L_m$. We present an example with $L_M=100$ and $l_m=10$,  which is the counterpart of Fig. \ref{delta}. After solving the gap and $\mu$ using summations, we plot the results in Fig. \ref{cmp} along with the results from the integrals. One can see that they are virtually identical, thereby justifying the approximation of replacing the summation over the angular momentum with integration. 

\section{Derivation of superfluid density}\label{app:SFder}
The superfluid density on a spherical shell can be deduced from the expression of the 2D planar case. We remark that the superfluid density is derived from linear response theory~\cite{FetterWalecka} instead of thermodynamics. Explicitly, $n_s$ can be extracted from the London equation $j_\mu=-\frac{n_s}{m}A_\mu$, where $j_\mu$ and $A_\mu$ denote the current and four-potential. From linear response theory, the current of a homogeneous system can be written as
\be
j_\mu(\vk,\omega)&=&-K_{\mu\nu}(\vk,\omega)A_\nu(\vk,\omega), \\
K_{\mu\nu}(\vk,\omega)&=&\frac{n}{m}\delta_{\mu\nu}-i\ep{[J_\mu(\vk,\omega),J_\nu(-\vk,-\omega)]}. \nonumber
\ee
Here $J_\mu(\vk,\omega)$ is the current operator and $\langle\cdots\rangle$ denotes the ensemble average. The current-current correlation function can be obtained by analytical continuation from the corresponding Matsubara formula. In the static and uniform limit with $\omega=0$ and $\vk\to0$, the result is simplified to
\be
\lim_{\vk\to0}-i\ep{[J_\mu(\vk,0),J_\nu(-\vk,0)]}=\frac{1}{m^2}\sum_\vk k^2\frac{\p f(E_\vk)}{\p E_\vk}\delta_{\mu\nu}.
\ee
After collecting all the above results, we find the $n_s$ of BCS theory on a 2D plane as
\be
n_s=n-\sum_k \epsilon_\vk\Big[-\frac{\p f(E_\vk)}{\p E_\vk}\Big].
\ee
To generalize the expression to the spherical case, we make the following replacements: 
\be
\sum_k\to \frac{1}{4\pi R^2}\int dl\,2(2l+1),\quad \epsilon_\vk\to\frac{l(l+1)}{2mR^2},
\ee 
Afterwards, we arrive at
$n_s=n-\frac{1}{4\pi R^2}\int dl\,2(2l+1)\frac{l(l+1)}{2mR^2}\Big[-\frac{d f(E_l)}{d E_l}\Big]$ as shown in the main text.
If the sphere is too small, the discreteness of the energy spectrum cannot be ignored. Then one has to replace the integral by an explicit summation of $l$.

\bibliographystyle{apsrev}

\end{document}